\documentclass[useAMS,usenatbib]{mn2e}

\usepackage{graphicx}
\usepackage{setspace}
\usepackage{natbib}
\usepackage{color}
\usepackage{amsmath,amssymb}
\usepackage{times}
\usepackage{aas_macros}

\title[New Milky Way satellite in Crater]{ATLAS lifts the Cup:
  Discovery of a New Milky Way satellite in Crater \thanks{Based on
    data products from observations made with ESO Telescopes at the La
    Silla Paranal Observatory under public survey programme ID
    programme 177.A-3011(A,B,C)} \thanks{Based in part on service
    observations made with the WHT operated on the island of La Palma
    by the Isaac Newton Group in the Spanish Observatorio del Roque de
    los Muchachos of the Instituto de Astrof\'isica de Canarias.}}

\author[Belokurov et al.]{V. Belokurov $^{1}$
  \thanks{E-mail:vasily@ast.cam.ac.uk}, M. J. Irwin $^{1}$,
  S. E. Koposov$^{1,2}$, N. W. Evans, $^{1}$,
  E. Gonzalez-Solares$^{1}$, \newauthor N. Metcalfe$^{3}$, T. Shanks
  $^{3}$ \\ $^{1}$Institute of Astronomy, Madingley Rd, Cambridge, CB3
  0HA, \\ $^{2}$Sternberg Astronomical Institute, Moscow State
  University, Universitetskiy pr. 13, Moscow 119991,
  Russia,\\ $^{3}$Department of Physics, Durham University, South
  Road, Durham DH1 3LE }

\begin{document}

\date{October 2013}
\pagerange{\pageref{firstpage}--\pageref{lastpage}} \pubyear{2013}

\maketitle

\label{firstpage}

\begin{abstract}
We announce the discovery of a new Galactic companion found in data
from the ESO VST ATLAS survey, and followed up with deep imaging on
the 4m William Herschel Telescope. The satellite is located in the
constellation of Crater (the Cup) at a distance of $\sim$ 170 kpc. Its
half-light radius is $r_h=30$ pc and its luminosity is $M_V=-5.5$. The
bulk of its stellar population is old and metal-poor. We would
probably have classified the newly discovered satellite as an extended
globular cluster were it not for the presence of a handful of Blue
Loop stars and a sparsely populated Red Clump. The existence of the
core helium burning population implies that star-formation occurred in
Crater perhaps as recently as 400 Myr ago. No globular cluster has
ever accomplished the feat of prolonging its star-formation by several
Gyrs. Therefore, if our hypothesis that the blue bright stars in
Crater are Blue Loop giants is correct, the new satellite should be
classified as a dwarf galaxy with unusual properties. Note that only
ten degrees to the North of Crater, two ultra-faint galaxies Leo IV
and V orbit the Galaxy at approximately the same distance. This hints
that all three satellites may once have been closely associated before
falling together into the Milky Way halo.
\end{abstract}


\section{Introduction}

By revealing 20 new satellites in the halo of the Milky Way
\citep{umadisc, willman1disc, cvndisc, boodisc, uma2disc, catsdisc,
  leotdisc, koposovdisc, boo2disc, leovdisc, segue2disc, boo3disc,
  piscdisc, balbinotdisc}, the Sloan Digital Sky Survey (SDSS) has
managed to blur the boundary between what only recently seemed two
entirely distinct types of objects: dwarf galaxies and star
clusters. As a result, we now appear to have ``galaxies'' with a total
luminosity smaller than that of a \textit{single} bright giant star
(e.g. Segue 1, Ursa Major II). The SDSS observations also wrought
havoc on intra-class nomenclature, giving us dwarf spheroidals with
properties of dwarf irregulars, i.e. plenty of gas and recent star
formation (e.g. Leo T), as well as distant halo globulars so
insignificant that if they lay ten times closer they would surely be
called open clusters (e.g. Koposov 1 and 2).

The art of satellite classification, while it may seem like idle
pettifoggery, is nonetheless of importance to our models of structure
formation.  By classifying the satellite, on the basis of all
available observational evidence, as a dwarf galaxy, we momentarily
gloss over the details of its individual formation and evolution, and
instead move on to modeling the population as a whole. We belive there
are more than twenty dwarf galaxies in the Milky Way environs and
predict that there are tens more dwarf galaxies waiting to be
discovered in the near future \citep[see e.g.][]{KoposovLF,
  Tollerud}. It is worth pointing out that the Cold Dark Matter (CDM)
paradigm remains the only theory that can easily produce large numbers
of dwarf satellites around spirals like the Milky Way or the M31. So
far the crude assumption of all dwarf galaxies being simple clones of
each other living in similar dark matter sub-halos has payed off and
the $\Lambda$CDM paradigm appears to have been largely vindicated
\citep[e.g.][]{KoposovLFmodel}. However, as the sample size grows, new
details emerge and may force us to reconsider this picture.

\begin{figure*}
  \centering
  \includegraphics[width=0.99\linewidth]{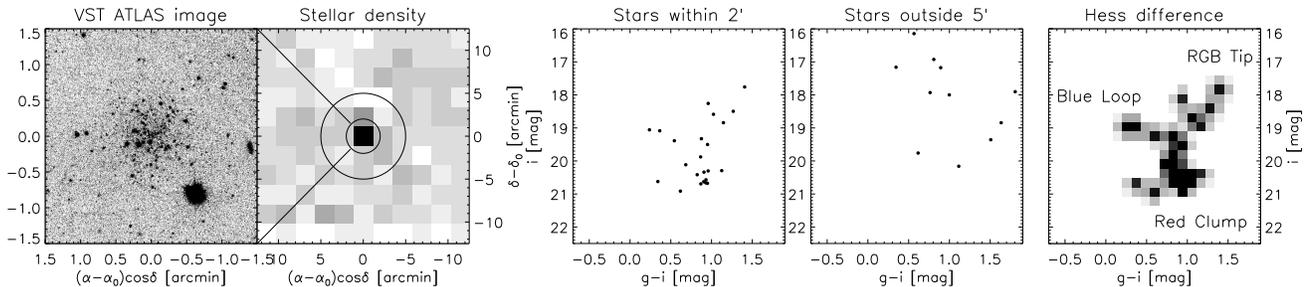}
  \caption[]{\small Discovery of the Crater satellite in VST ATLAS
    survey data. \textit{First:} 3x3 arcminute ATLAS $r$-band image
    cutout of the area around the centre of the
    satellite. \textit{Second:} Stellar density map of a 25x25
    arcminute region centered on Crater and smoothed using a Gaussian
    kernel with FWHM of 2 arcmin.  Darker pixels have enhanced
    density.  The smaller circle marks the region used to create the
    CMD of the satellite stars. Bigger circle denotes the exclusion
    zone used when creating the CMD of the Galactic
    foreground. \textit{Third:} CMD of the 2 arcmin radius region
    around the centre of Crater.  Several coherent features including
    a Red Giant Branch and Red Clump are visible. Also note several
    stars bluer and brighter than the Red Clump. These are likely Blue
    Loop candidate stars indicating recent ($< 1$ Gyr) star formation
    activity in the satellite. \textit{Fourth:} CMD of the Galactic
    foreground created by selecting stars that lie outside the larger
    circle marked in the Second panel and covering the same area as
    that within the small circle. It appears that the satellite CMD is
    largely unaffected by Galactic contamination. \textit{Fifth:} Hess
    difference of the CMD density of stars inside the small circle
    shown and stars outside the large circle.}
   \label{fig:disc}
\end{figure*}

The distribution of known satellites around the Milky Way is
anisotropic, though this is partly a consequence of selection effects.
It has been claimed that the Galactic satellites form a vast disc-like
structure about 40 kpc thick and 400 kpc in diameter, and that this is
inconsistent with $\Lambda$CDM~\citep[e.g.,][]{Kr12}. In fact, as the
recent discoveries have been made using the SDSS, a congregation of
satellites in the vicinity of the North Galactic Cap is only to be
expected~\citep{Be13}. Even then, anisotropic distributions of
satellite galaxies do occur naturally within the $\Lambda$CDM
framework. For example, using high resolution hydrodynamical
simulations, \citet{De11} found that roughly 20 \% of satellite
systems exhibit a polar alignment, reminiscent of the known satellites
of the Milky Way galaxy. In 10\% of these systems there was evidence
of satellites lying in rotationally supported discs, whose origin can
be traced back to group infall. Significant fractions of satellites may
be accreted from a similar direction in groups or in loosely bound
associations and this can lead to preferred planes in the satellite
distributions~\citep[e.g.,][]{Li08,Do08}. There is ample evidence of
such associations in the satellites around the Milky Way -- such as
Leo IV and Leo V ~\citep{Jo10} -- and M31 -- such as NGC 147 and NGC
185 and possibly Cass II~\citep{Wa13,Fa13}.

Understanding how dwarf galaxies and clusters evolve in relation to
their environment remains a challenge. These objects can occupy both
low and high density environments and are subject to both internal
processes (mass segregation, evaporation and ejection of stars, bursts
of star formation) and external effects (disruption by Galactic tides,
disk and bulge shocking, ram pressure stripping of gas by the ISM).
There is also growing evidence for interactions and encounters of
satellites with each other, for example, in the tidal stream within
And II~\citep{Am14} and the apparent shells around
Fornax~\citep{Co04,Am12}.  The variety of initial conditions near the
time of formation, together with the diversity of subsequent
evolutionary effects in a range of different environments, is capable
of generating a medley of objects with luminosities and sizes of the
present day cluster and dwarf galaxies populations. It is therefore
unsurprising that as the Milky Way halo is mapped out we are finding
an increasing number of ambiguous objects that do no fit tidily into
the once clear-cut categories of clusters and dwarf galaxies.

In this paper, we describe how applying the overdensity search
algorithms perfected on the SDSS datasets to the catalogues supplied
by the VST ATLAS survey has uncovered a new satellite in the
constellation of Crater. Although Crater has a size close to that of
globular clusters, we believe that the satellite has had an extended
star-formation history and therefore should be classified as a galaxy
\citep[see e.g.][]{whatisgalaxy}. Therefore, following convention, it
is named after the constellation in which it
resides. Section~\ref{sec:data} gives the particulars of the ATLAS
data and of the follow-up imaging we have acquired with the 4m
WHT. Section~\ref{sec:prop} describes how the basic properties of
Crater were estimated. Section~\ref{sec:conclusions} provides our
Discussion and Conclusions.

\begin{figure}
  \centering
  \includegraphics[width=0.97\linewidth]{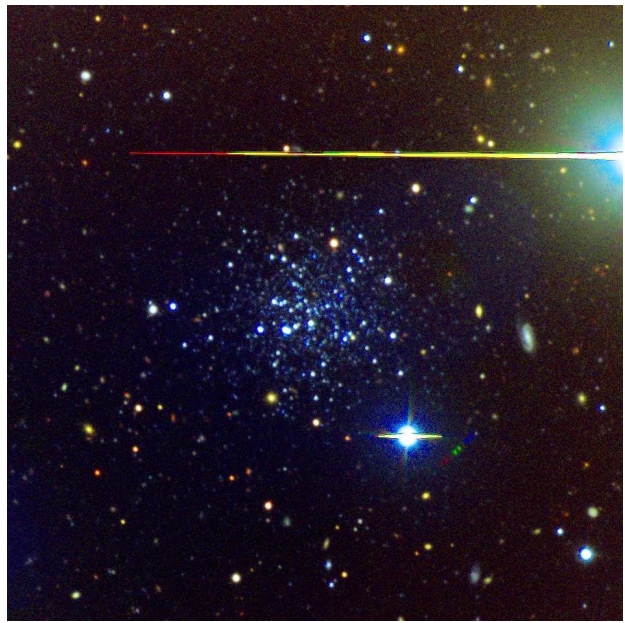}
  \caption[]{\small False-color WHT ACAM image cutout of an area 4x4
    arcminutes centered on Crater. The $i$-band frame is used for the
    Red channel, $r$-band for the Green and $g$-band for the Blue. The
    image reveals the dense central parts of Crater dominated by faint
    MSTO and SGB stars. Several bright stars are clearly visible,
    these are the giants on the RGB and in the Red Clump. Amongst
    these are 3 or 4 bright Blue Loop giant candidates. A sprinkle of
    faint and very blue stars is also noticeable. These are likely to
    be either young MS stars or Blue Stragglers.}
   \label{fig:image}
\end{figure}
\begin{figure*}
  \centering
  \includegraphics[width=0.95\linewidth]{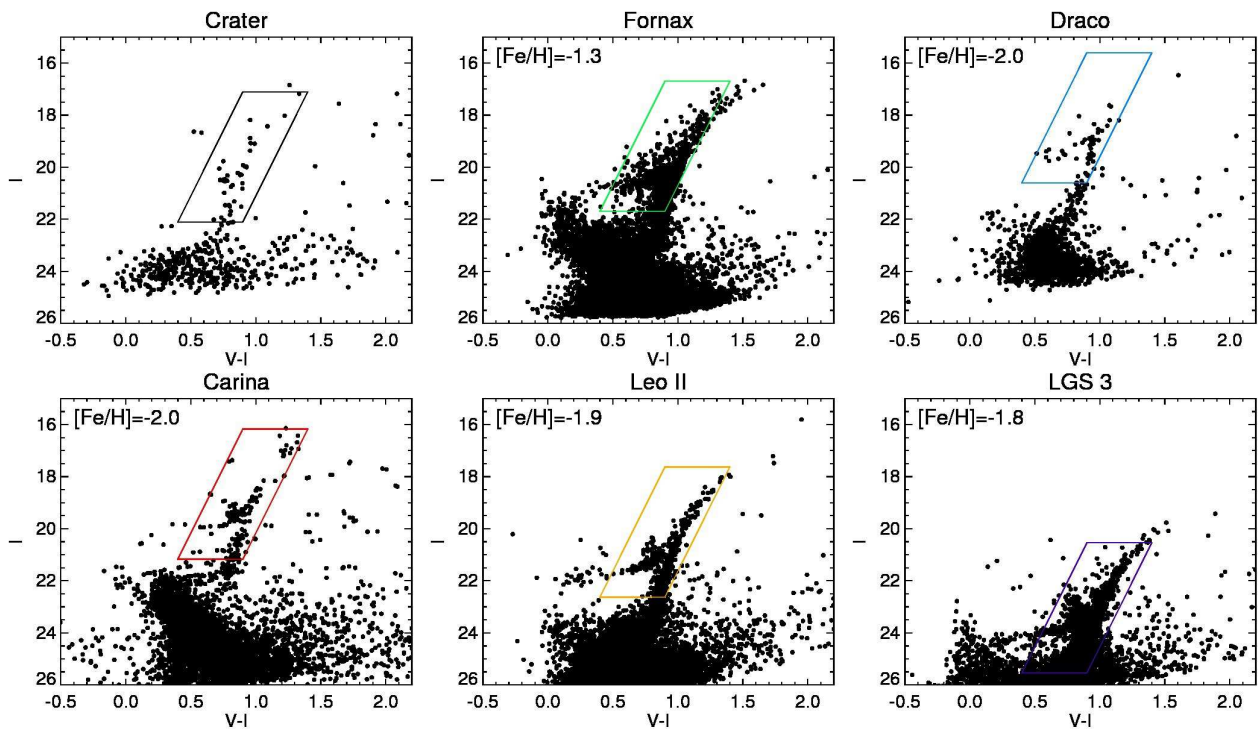}
  \caption[]{\small Red Clump morphology. The CMD features around the
    RC region in Crater (\textit{Top Left}) are contrasted to those in
    5 classical dwarf galaxies observed with the HST
    \citep{HoltzmanHST}. The polygon mask shows the boundary used to
    select the RC/RHB and RGB stars for the Luminosity Function
    comparison in Figure~\ref{fig:lf}. The RC/RHB region of Crater
    most closely resembles that of Carina, the main differences being
    the lack of an obvious Horizontal Branch and the presence of a
    small but visible Blue Loop extension. \textit{Top Center:}
    Fornax, at $(m-M)=20.7$ and E(B-V)=0.025. \textit{Top Right:}
    Draco, at $(m-M)=19.6$ and E(B-V)=0.03. \textit{Bottom Left:}
    Carina, at $(m-M)=20.17$ and E(B-V)=0.06. \textit{Bottom Center:}
    Leo II, at $(m-M)=21.63$ and E(B-V)=0.017. \textit{Bottom Right:}
    LGS 3, at $(m-M)=24.54$ and E(B-V)=0.04. The dwarf photometry is
    extinction corrected.}
   \label{fig:redclump}
\end{figure*}
\begin{figure}
  \centering
  \includegraphics[width=0.99\linewidth]{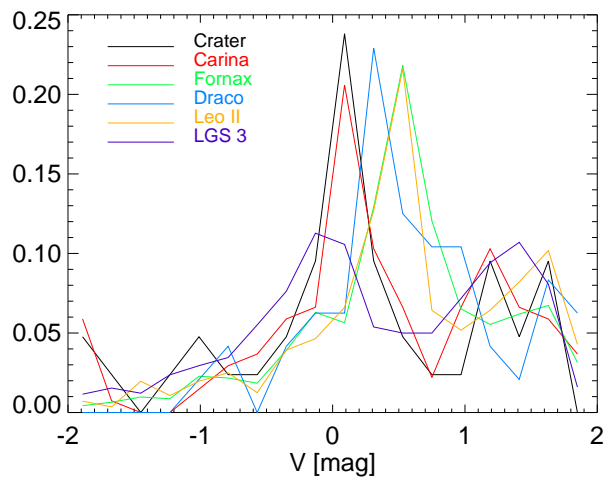}
  \caption[]{\small Luminosity Functions of stars selected using the
    polygon CMD masks in 6 satellites as shown in
    Figure~\ref{fig:redclump}, given the distance moduli and the
    extinction values recorded in Figure~\ref{fig:redclump}
    caption. The difference in RC/RHB morphology is reflected in the
    difference of LF shapes and, most importantly, in different
    absolute magnitudes of the LF peak. Crater's stars are offset to
    $(m-M)=21.1$ to match the peak of Carina's LF.}
   \label{fig:lf}
\end{figure}

\section{Data, Discovery and Follow-up}
\label{sec:data}

\subsection{VST ATLAS}

\begin{figure}
  \centering
  \includegraphics[width=0.99\linewidth]{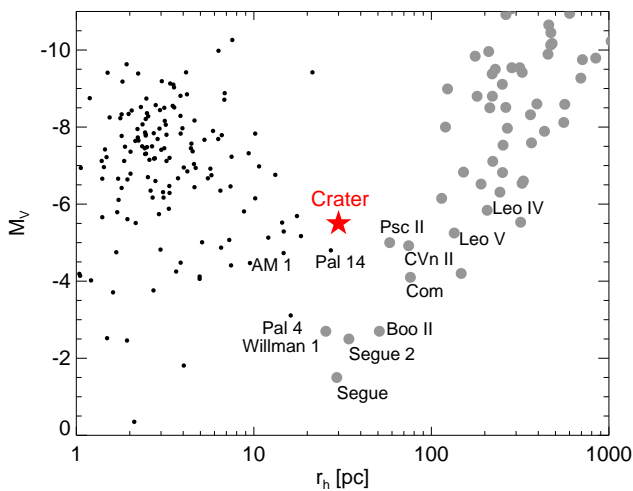}
  \caption[]{\small Luminosity $M_V$ versus half-light radius $r_h$
    for Galactic globular clusters (black dots) and Milky Way and M31
    satellites (grey circles). The red star marks the location of
    Crater. The new satellite appears to have more in common with the
    Galactic globular cluster population, even though it appears to be
    the largest of all known Milky Way globulars. Contrasted with the
    known ultra-faint satellites, Crater is either too small for the
    given luminosity, c.f. Pisces II, Leo V or too luminous for a
    given size, c.f. Willman 1, Segue.}
   \label{fig:rhmv}
\end{figure}
\begin{figure*}
  \centering
  \includegraphics[width=0.99\linewidth]{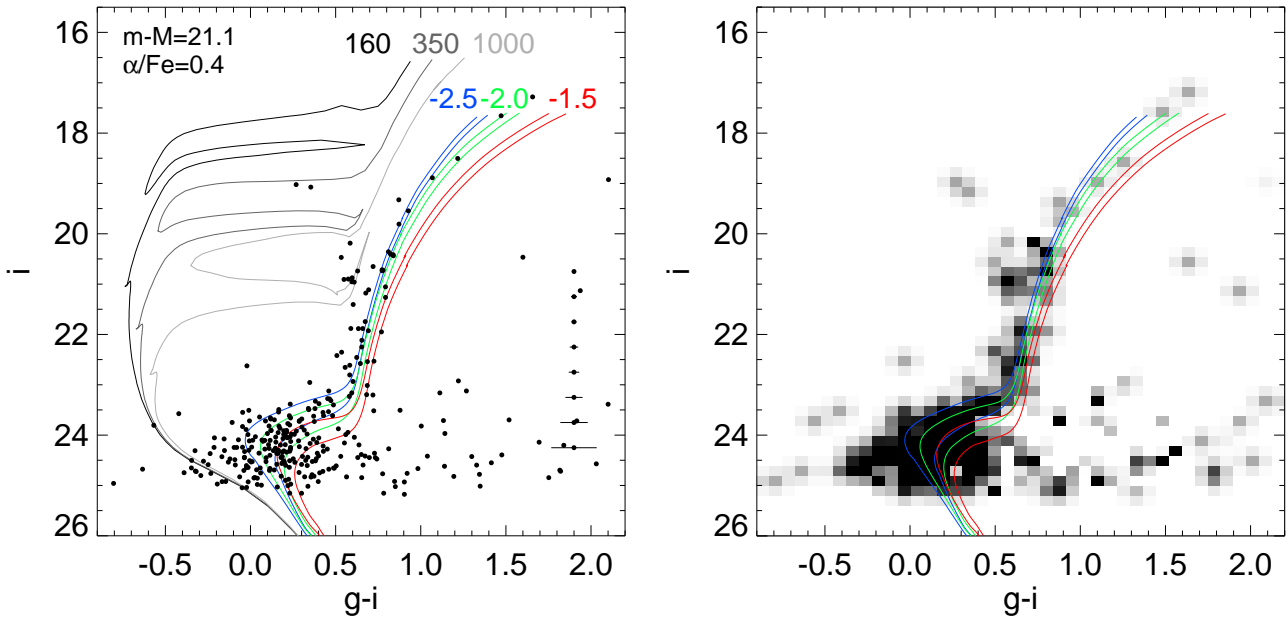}
  \caption[]{\small Colour magnitude diagram of the Crater stars
    within $2r_{\rm h}$ of the center. \textit{Left:} All WHT ACAM
    sample A stars inside two half-light radii. For comparison, 7 and
    10 Gyr and [$\alpha/Fe]=0.4$ Dartmouth isochrones with
    metallicities [Fe/H]=$-$2.5 (blue), [Fe/H]=$-$2.0 (green) and
    [Fe/H]=$-$1.5 (red) are over-plotted. Also shown are 160 Myr
    (black), 350 Myr (dark grey) and 1 Gyr (light grey) Padova
    isochrones. If the RC distance calibration is trustworthy, then
    the bulk of Crater's stellar population is old (between 7 and 10
    Gyr) and metal-poor ($-$2.5 $<$ [Fe/H] $< -$2.0). However, there are
    three peculiar stars, two directly above the blue side of the Red
    Clump and one at $g-i \sim 0.3$ and $i\sim 19$ that can be
    interpreted as Blue Loop giants. If this interpretation is
    correct, then according to the Padova isochrones, Crater has
    experienced a small amount of star-formation around 1 Gyr ago and
    perhaps as recently as 350 Myr ago. \textit{Right:} Hess
    difference between the CMD densities of the stars inside $r_h$ and
    the stars outside $4.5 r_h$. Note the tightness of the sub-giant
    region.}
   \label{fig:cmd}
\end{figure*}

ATLAS is one of the three ESO public surveys currently being carried
out at Paranal in Chile with the 2.6m VLT Survey Telescope
(VST). ATLAS aims to cover a wide area of several thousand square
degrees in the Southern celestial hemisphere in 5 photometric bands,
$ugriz$, to depths comparable to those reached by the Sloan Digital
Sky Survey (SDSS) in the North. Compared to the SDSS, ATLAS images
have finer pixel sampling and somewhat longer exposures, while the
median seeing at Paranal is slightly better than that at Apache
Point. We note that the resulting default catalogued photometry
produced by the Cambridge Astronomical Survey Unit (CASU) is in the
Vega photometric system rather than AB. The median limiting magnitudes
in each of the five bands corresponding to 5$\sigma$ source detection
limits are approximately $21.0, 23.1, 22.4, 21.4, 20.2$. The details
of the VST image processing and the catalogue assembly can be found in
\citet{atlasstream}. For the analysis presented in this paper, we have
transformed ATLAS Vega magnitudes to APASS AB system by adding -0.031,
0.123 and 0.412 to the $g$, $r$ and $i$-band magnitudes
respectively. Subsequently all of these magnitudes are corrected for
Galactic extinction using the dust maps of \citet{Schlegel}.

\subsection{Discovery}

Crater was discovered in October 2013 during a systematic search for
stellar overdensities using the entirety of the ATLAS data processed
to that date ($\sim 2700$ square degrees). The satellite stood out as
the most significant candidate detection, as illustrated by the second
panel of Figure~\ref{fig:disc} showing the stellar density
distribution around the center of Crater. In this map, the central
pixel corresponds to an overdensity in excess of 11$\sigma$ over the
Galactic foreground. The identification of the stellar clump as a
genuine Milky Way satellite was straightforward enough as the object
was actually visible on the ATLAS frames\footnote{In fact, Crater is
  visible on the Digitized Sky Survey images as well, similarly to Leo
  T} as evidenced by the first panel of Figure~\ref{fig:disc}. The
colour magnitude diagram (CMD) of all ATLAS stars within 2 arcminute
radius around the detected object is presented in the third panel of
the Figure. There are three main features readily discernible in the
CMD as well as in the accompanying Hess difference diagram (shown in
the fifth panel of the Figure). They are the Red Giant Branch (RGB)
with $18 < i < 21$, the Red Clump (RC) at $i \approx 20.5$ and the
likely Blue Loop (BL) giants, especially the two brightest stars with
$g-i < 0.5$ and $i \approx 19$. While the identification of these
stellar populations seems somewhat tenuous given the scarcity of the
ATLAS CMD, it is supported by the deeper follow-up imaging, as
described below.

\subsection{Follow-up imaging with the WHT}

The follow-up WHT imaging data were taken using the Cassegrain
instrument ACAM in imaging mode.  The observations were executed as
part of a service programme on the night of 11th February 2014 and
delivered 3x600s dithered exposures in the $g$, $r$ and $i$-bands with
seeing of around 1 arcsec. The effective field-of-view of ACAM is some
8 arcmin in diameter with a pixel scale of 0.25 arcsec/pixel.

\begin{figure}
  \centering
  \includegraphics[width=0.98\linewidth]{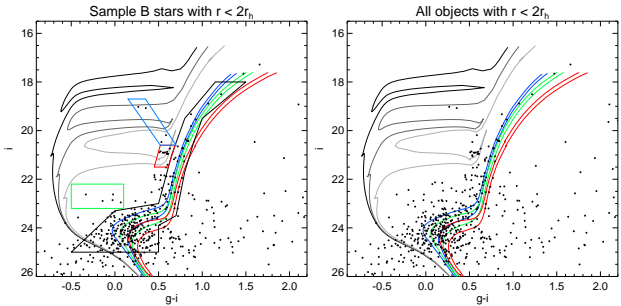}
  \caption[]{\small CMD evolution with morphological
    classification. \textit{Left:} CMD of objects inside $2 r_h$ in
    Sample B. Compared to the sample of stars with cleanest photometry
    (Sample A) shown in Figure~\ref{fig:cmd}, the CMD looks almost
    identical at magnitudes brighter than that of the Red
    Clump. However, faintwards of the RC there are noticeable
    differences. Note several faint blue stars above the MSTO. These
    are either young MS stars or Blue Stragglers. Also shown are CMD
    masks used to select the MSTO/RGB (black), Red Clump (red), Blue
    Loop (blue) and young MS (green) populations. The distribution of
    the selected stars on the sky is shown in
    Figure~\ref{fig:den}. \textit{Right:} CMD of all objects within $2
    r_h$. The region directly above the MSTO is now densely
    populated. Most likely, the photometry of these stars have been
    affected by blending, more precisely, given their colors and
    magnitudes by blending of YMS/BS stars and SGB/RGB populations.}
   \label{fig:cmd2}
\end{figure}

A series of bias and twilight flat frames taken on the same night were
used to bias-correct, trim and flatfield the data. Fringing, even on
the $i$-band images, was negligible. Figure~\ref{fig:image} shows a
false-color image cutout created using the ACAM $i$, $r$ and $g$-band
frames for the Red, Green and Blue channels respectively. This 4x4
arcminute fragment reveals the dense central parts of the
satellite. Concentrated around the middle are a handful of bright
stars (saturated in this false-color image and therefore appearing
white). As the following Section makes clear, these are the Red Giant
Branch (RGB) stars, Red Clump giants and possible Blue Loop
giants. The fainter white stars are the Main Sequence Turn Off (MSTO)
dwarfs. Also visible in this image are faint and very blue stars which
must be either a young MS or the Blue Straggler population.

Object catalogues were generated from the processed images to refine
the World Coordinate System transformation. Each set of three
individual dithered exposures per band were then stacked with bad
pixel and cosmic ray rejection and further catalogues from the
combined images generated.  These catalogues were then morphologically
classified and the photometry cross-calibrated with respect to the VST
ATLAS $g$, $r$ and $i$ band data, yielding the colour-magnitude
diagram shown in Figures~\ref{fig:redclump}, \ref{fig:cmd} and
\ref{fig:cmd2}. While overall, the innards of Crater appear resolved
in the ACAM images, the resulting photometry does suffer from blending
problems to some extent. Therefore, in the following analysis we use
two versions of the WHT stellar catalogues. First, the most stringent
sample (sample A) is created by requiring that each object is
classified as stellar in both $r$ and $i$ frames. This produces clean
photometry across the face of the satellite, but leads to a pronounced
depletion in the very core. The total is 1271 star-like objects across
the ACAM frame and 324 within $2 r_{h}$. We use this catalogue for the
colour magnitude diagram studies. Alternatively, if a more relaxed
condition of being detected in one of the three of the $g$, $r$ or $i$
bands is applied, the total of 1964 (496 inside 2 $r_{h}$) objects
classified as stars is recovered. We use this sample (sample B) to
study the density distribution and to measure the structural
parameters of Crater, including its luminosity. Note that the ratios
of the objects in each of the two catalogues also reflect the relative
decrease in the number of genuine background galaxies in the more
stringent sample. If no morphological classification is enforced,
there are 2228 objects in total across the ACAM frame, with 550
objects inside $2 r_h$.

\section{Properties of Crater}
\label{sec:prop}

\subsection{Distance}
\label{sec:distance}

\begin{figure}
  \centering
  \includegraphics[width=0.99\linewidth]{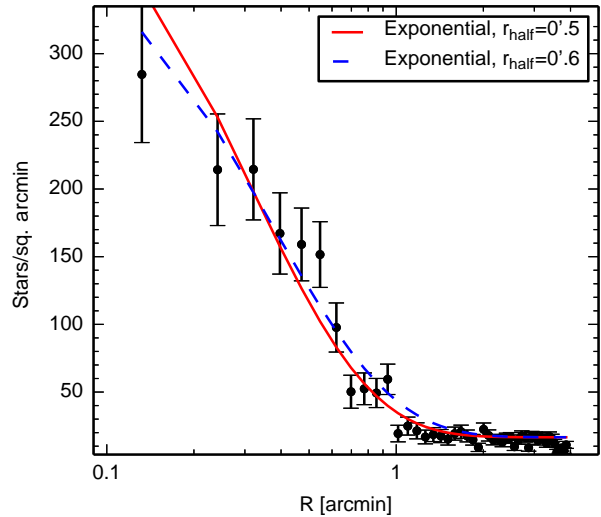}
  \caption[]{\small Surface brightness profile of stars in the ACAM
    sample B with $i<24.5$ (black). Overlayed are exponential density
    models with half-light radius of $r_h=0 \farcm5$ (red, solid) and
    $r_h=0\farcm6$ (dashed, blue).}
   \label{fig:profile}
\end{figure}
\begin{figure}
  \centering
  \includegraphics[width=0.99\linewidth]{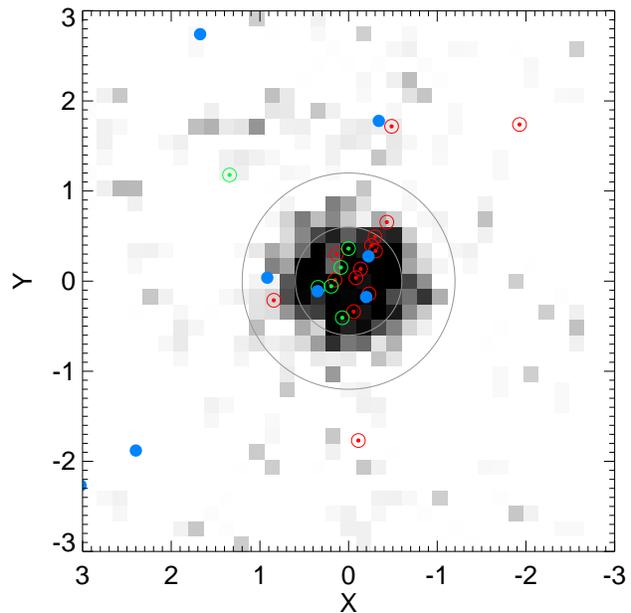}
  \caption[]{\small Distribution of Crater stars on the sky for three
    populations selected using the CMD masks shown in the bottom right
    panel of Figure~\ref{fig:cmd2}. The greyscale reflects the density
    distribution of the MSTO stars. Red circles mark the locations of
    the Red Clump giants, blue filled circles are Blue Loop stars, and
    green circles are young MS candidates. Large grey circles give the
    $r_h$ and $2 r_h$ boundaries.}
   \label{fig:den}
\end{figure}

The best distance indicator available amongst the stellar populations
in Galactic satellites are the Blue Horizontal Stars, either stable or
pulsating, i.e. RR Lyrae. Unfortunately, as Section~\ref{sec:cmd}
demonstrates, these are not present in Crater. However, the newly
discovered satellite possesses a well-defined Red Clump. Therefore we
endeavour to estimate Crater's heliocentric distance by calibrating
the average intrinsic luminosity of its Red Clump stars with the
extant photometry of similar stellar populations in the dwarf galaxies
with known distances. To this end we take advantage of the publicly
available treasure trove of broad band photometry of the Local Group
dwarf galaxies observed with the Hubble Space Telescope (HST)
described in \citet{HoltzmanHST}. We chose five dwarfs with prominent
Red Clump and/or Red Horizontal Branch (RC/RHB) populations, namely
Fornax, Draco, Carina, Leo II and LGS 3. To compare the WHT data to
the HST photometry, we transform the APASS AB $g$ and $i$ magnitudes
of Crater stars to $V$ and $I$ magnitudes using the standard equations
of
Lupton~\footnote{http://www.sdss.org/dr7/algorithms/sdssUBVRITransform.html}.

To select the RGB and the RC/RHB stars in the above mentioned
galaxies, we apply the color-magnitude mask shown in
Figure~\ref{fig:redclump} offset to the appropriate distance modulus
(listed for each object in the Figure caption). The color of the
polygon used for selection corresponds to the color of the luminosity
function (LF) curve shown for each dwarf galaxy in
Figure~\ref{fig:lf}. As evident from Figure~\ref{fig:redclump}, the
morphology of the RC/RHB is subtly (but noticeably) different in each
case. This is reflected in the shape of the LF in
Figure~\ref{fig:lf}. For example, Draco, Fornax and Leo II have a
property in common: a substantial RHB, pushing the peak of the RC/RHB
region faintwards. This is to be compared with LGS 3 which exhibits a
noticeable stump of Blue Loop stars extending from RC/RHB to brighter
magnitudes at bluer $V-I$. As a result, in LGS 3 the RC/RHB luminosity
function peaks some 0.7 mag brighter as compared to Leo II and
Fornax. Morphologically, the RC/RHB of Carina appears to be the
closest match to Crater's. Carina's RC/RHB luminosity function reaches
a maximum at $M_{V}=0.1$, which gives $(m-M)=21.1$, or $\sim$ 170 kpc
for the distance of Crater.

\begin{figure}
  \centering
  \includegraphics[width=0.98\linewidth]{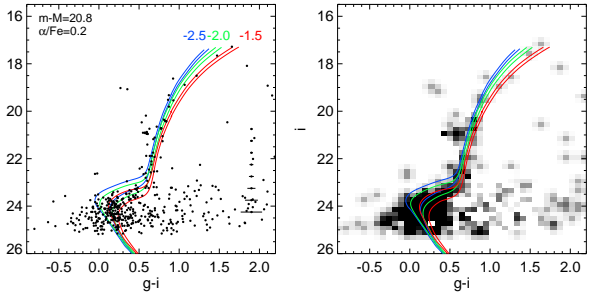}
  \caption[]{\small See Figure~\ref{fig:cmd} for
    detail. \textit{Left:} CMD of star-like objects inside $3 r_h$ in
    Sample A. Compared to Figure~\ref{fig:cmd}, a lower (more
    cluster-like) distance modulus is assumed $(m-M)=20.8$ as well as
    a lower enhancement for the Dartmouth isochrones
    $[\alpha/Fe]=0.2$. The change in distance and in the light
    elements budget results in a slightly better agreement with the
    metal-rich isochrones. However, the RGB of a population with
    [Fe/H]=$-$1.5 remains too red compared to Crater's.}
   \label{fig:cmd3}
\end{figure}

Alternatively, it is possible to gauge the distance to the new
satellite by taking advantage of the gently sloping correlation
between the absolute magnitude of the Horizontal Branch and the
metallicity established for the Galactic Globular Clusters. As
apparent from comparison of the panels in Figure~\ref{fig:redclump},
the Crater Red Giant Branch is bluer and steeper compared to most of
the dwarf spheroidal populations displayed. Therefore, as an initial
estimate, the plausible metallicity range could be as low as [Fe/H] $<
-$1.8. According to the study of \citet{DotterHB}, for a globular
cluster of low metallicity $M^{HB}_{V}\sim 0.4$ which would bring
Crater slightly closer to the Sun at $(m-M)=20.8$, or 145 kpc. As
Section~\ref{sec:cmd} details, Crater's CMD does indeed appear
consistent with an old metal-poor stellar population. Note, however,
that no globular cluster with [Fe/H]$< -$1 in the sample analyzed by
\citet{DotterHB} boasts as red a Horizontal Branch as that found in
Crater.

\subsection{Size and Luminosity}
\label{sec:sizelum}

\begin{figure*}
  \centering
  \includegraphics[width=0.98\linewidth]{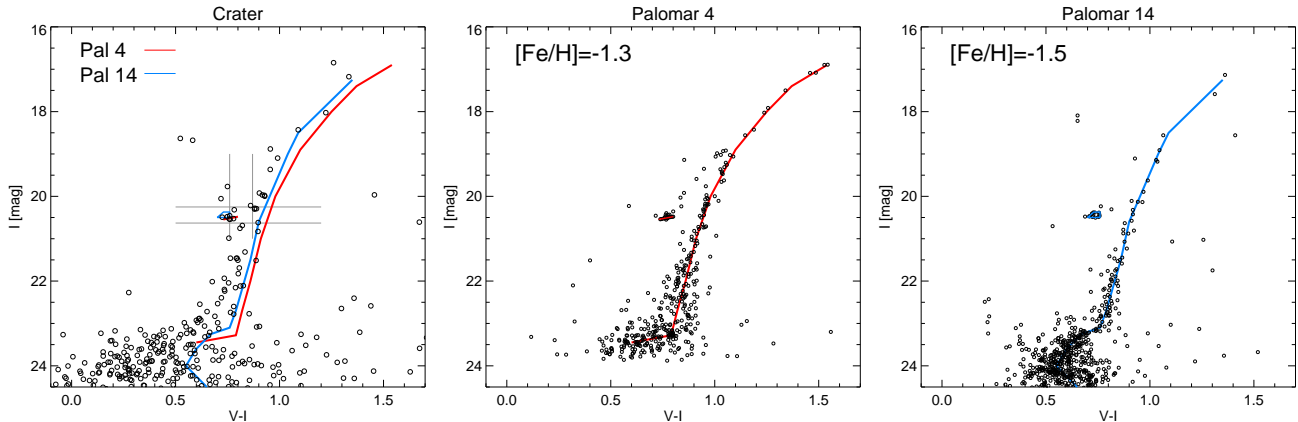}
  \caption[]{\small Crater's CMD compared to two extreme outer halo
    Globular Clusters Pal 4 (at $(m-M)=20.21$) and Pal 14 (at
    $(m-M)=19.54$). The Pal 4 and Pal 14 distances are from
    \citet{DotterHB}. \textit{Left:} CMD of star-like objects inside $2
    r_h$ in Sample A. Blue (red) line shows the ridgeline of Pal 14
    (Pal 4) offset to $(m-M)=20.8$. We conclude that Crater must be
    more metal-poor than [Fe/H]=$-$1.5 as both ridgeline RGBs are
    clearly redder. Note that the apparent magnitudes of the RHBs of
    both Pal 4 and Pal 14 match very well the location of the RC/RHB
    in Crater, lending support to our distance determination. Grey
    horizontal lines show the stars selected for the $\Delta(V-I)$
    measurement. Grey vertical lines show the median $V-I$ colors of the
    RC/RHB and RGB stars selected, resulting in the value of
    $\Delta(V-I)=0.11$. \textit{Middle:} CMD of Pal 4 stars from
    \citet{paldata} offset to the distance $(m-M)=20.8$. The red line
    shows the approximate ridgeline. \textit{Right:} CMD of Pal 14
    stars from \citet{paldata} offset to the distance
    $(m-M)=20.8$. The blue line shows the approximate ridgeline. }
   \label{fig:cmd4}
\end{figure*}

The structural properties of Crater are measured through maximum
likelihood (ML) modeling \citep[see e.g.][]{MartinML} of the positions of
stars with $g-i < 1.6$ selected from the ACAM sample B. The results of
the ML analysis are reported in Table 1. While the nominal error on the
half-light radius is rather small, given that the central parts of
Crater suffer from some amount of blending, it could perhaps be
slightly overestimated. Note however that switching from sample B to
A, i.e. losing as much as $\sim 35\%$ of star-like objects within $2
r_{\rm h}$ leads only to a minor increase in half-light radius from
0.6 to 0.7 arcminutes. We have investigated further the effects
completeness might have on the stellar density profile
parameterization. If the sample B is limited to stars brighter than
$i=24.5$, i.e. objects less affected by blending, then the structural
parameters of Crater evolve as follows. For the brighter sample, we
get $r_h$ between $0\farcm 5$ an $0\farcm 55$ and the ellipticity of
$0.2\pm 0.05$ at the position angle of $110^{\circ}$. As
Figure~\ref{fig:profile} illustrates, the exponential profile with
half-light radius between $r_h=0\farcm5$ and $r_h=0 \farcm 6$ is a
reasonable fit to the ACAM data. Note that at a distance of 170 kpc,
$0\farcm6$ corresponds to 30 pc (or 25 pc at 145 kpc). Depending on
the tracer sample used, the density flattening is either consistent
with zero (all stars) or very slight elliptical (bright cut). For
further analysis, we chose to assume the ellipticity of Crater to be
zero. The question of the satellite's true shape can only be resolved
with a higher resolution dataset.

To estimate the total luminosity of the satellite, we sum up the flux
of all sample B stars inside $3 r_h$ that fall within a generous
colour-magnitude mask enclosing the MSTO, the RGB, the RC and the
BL. The total is $V_{\rm tot}= 15.8$, for a miniscule foreground
contamination of 18 mag. Therefore, assuming the distance modulus of
$(m-M)=21.1$, the total luminosity of Crater down to the MSTO is $M_V
= -5.2$. By integrating the LF of the old and metal poor globular
cluster M92 faintwards of the turn-off, we obtain the correction of
$\sim 0.3$ mag in missing flux, giving the final estimate of the total
luminosity $M_V=-5.5$. This is likely to be the lower bound as we have
not accounted for the stars missing from sample B due to blending or
any additional populations fainter than the detected MSTO.

As Figure~\ref{fig:rhmv} illustrates, at $M_V=-5.5$ and $r_h = 30$ pc,
Crater looks more like a puffed up globular cluster rather than a
dwarf spheroidal galaxy. In the Milky Way there are examples of faint
and extended clusters that might superficially resemble Crater, for
example, Palomar 14, which is only $\sim 1$ mag less luminous. The
only known galaxies that have some properties in common with Crater
are the so-called Ultra Faint Dwarfs (UFDs). These, however are either
much bigger at the same luminosity like Pisces II or Leo V, or much
fainter at the same size like Willman 1 or Segue. Of course, the total
luminosity and the extent of the stellar distribution alone do not
determine the satellite's nature. While many previously existing
boundaries have been blurred with the discovery of the ultra-faint
dwarfs, an extended star-formation history generally counts as a
feature inherent to galaxies and not star clusters. That being the
case, it is worth taking a closer look at the stellar populations of
Crater.

\begin{table}
\caption{Properties of the Crater satellite}
\centering
\begin{tabular}{lc}
\hline
Property & \null \\
\hline
Coordinates (J2000) & $\alpha = 11:36:15.8  $, $\delta = -10:52:40$ \\
Coordinates (J2000) & $\alpha = 174.066^{\circ}  $, $\delta = -10.8777^{\circ}$ \\
Coordinates (Galactic) & $\ell = 274.8^\circ$, $b= 47.8^\circ$ \\
Ellipticity & $0.08 \pm 0.08$\\
$r_h$ (Exponential) & $0\farcm 6 \pm 0\farcm 05$\\
$V_{\rm tot}$ & $15\fm 8 \pm 0\fm 2$\\
(m-M)$_0$ & 20\fm 8 - $21\fm 1$\\
M$_{\rm tot,V}$ & $-5\fm 5 \pm 0\fm 5$\\
\hline
\end{tabular}
\end{table}

\subsection{Stellar Populations}
\label{sec:cmd}

Figure~\ref{fig:cmd} gives the positions on the CMD of all stars in
sample A located within two half-light radii from the Crater's
center. The ACAM photometry reaches as faint as $i \sim 25$, the depth
sufficient to detect the Sub-Giant Branch (SGB) and the Main Sequence
Turn-off at $\sim$ 170 kpc. As the Hess difference (shown in the right
panel of the Figure) emphasizes, the SGB is very tight. To put
constraints on the metallicity and the age of the system, three sets
of Dartmouth isochrones \citep{Dartmouth} with different metallicities
are also over-plotted. For each of the three values of [Fe/H] ($-$2.5,
$-$2 and $-$1.5 corresponding to blue, green and red curves), two ages
are shown, namely 7 and 10 Gyr. If the RC distance calibration is
correct, then the bulk of Crater's stellar content has to be
metal-poor, i.e. somewhere between [Fe/H]$=-$2.5 and [Fe/H]$=-$2.0 and
old, i.e. with an age between 7 and 10 Gyr.

Figure~\ref{fig:cmd} also reveals 4 stars, bluer and brighter than the
Red Clump. Two of these are directly above the RC location at $i\sim
20.5$, and the two bluest ones are at $i\sim 19$. We interpret these
bright and blue stars as belonging to the core helium burning Blue
Loop population. As the name suggests, the colour and the luminosity
of the BL stars vary markedly during their lifetime. However, their
peak luminosities are robust indicators of age and as such BL stars
have been used successfully to map star-formation histories in dwarf
galaxies~\citep[e.g.,][]{To98}. According to the Padova isochrones
\citep{Girardi} over-plotted in the left panel of
Figure~\ref{fig:cmd}, the age of the two faint BL stars is of order of
1 Gyr while the brightest one is somewhat younger at $\sim$ 400
Myr. Figure~\ref{fig:den} shows the distribution of the BL candidates
in and around Crater. According to the Figure, there is very little
Galactic contamination from stars with colors and magnitudes typical
of the Crater's BL and RC populations.

If it were not for the presence of the handful of possible Blue Loop
giants, classifying the new satellite would be easy as it has more
features in common with the currently known Galactic globular clusters
than with classical or ultra-faint dwarfs.  Spectroscopic follow-up is
needed to firmly establish whether the Blue Loop candidates are indeed
members of Crater and not mere Galactic foreground
contaminants. Surprisingly, Figure~\ref{fig:cmd} does not include the
low-mass MS counterparts to the more massive BL stars discussed
above. Figure~\ref{fig:cmd2} helps to explain why. To be included into
Sample A, objects must be classified as stellar in both $r$ and $i$
bands\footnote{Unfortunately the seeing in the $g$ band data was
  substantially worse compared to $r$ and $i$}. However, for a young
MS star, most of the flux would be detected in the $g$ band. The left
panel of the Figure shows the CMD of all Sample B objects
(i.e. objects classified as stars in $r$, $i$ or $g$ bands) within $2
r_h$. A group of several blue stars is clearly visible above the MSTO
(also see Figure~\ref{fig:den}). The right panel of the Figure shows
the CMD for all objects inside $2r_h$ independent of their
morphological classification. The region above the MSTO is now more
densely populated. Given the colors and the magnitudes of these
objects, they could be young MS (or BS) stars blended with SGB/RGB
population. Therefore, it seems that there could be a substantial
young MS population in Crater corresponding to the handful of more
massive BL giants. Considering the blending problems affecting
photometry in the central parts, higher resolution imaging is required
to further probe the CMD of the satellite.

\begin{figure}
  \centering
  \includegraphics[width=0.98\linewidth]{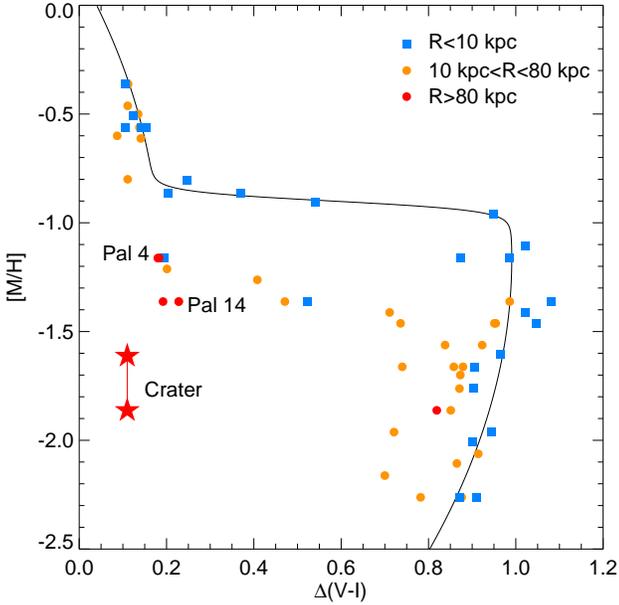}
  \caption[]{\small Morphology of the Horizontal Branch as a function
    of metallicity in Galactic Globular Clusters. This shows
    metallicity [M/H] =
    [Fe/H]+$\log_{10}$(0.638$\times$10$^{[\alpha/{\rm Fe}]}$ + 0.362)
    as a function of the difference in median color between the HB and
    the RGB $\Delta(V-I)$. Blue squares mark the locations of the
    inner globulars, while orange circles indicate the outer GCs. Red
    circles show the properties of the most distant globular clusters
    including Pal 4 and 14. The solid black line is a fit from
    \citet{DotterHB}. Note that the HB morphology in Crater is rather
    unusual compared to the rest of the Milky Way population: it is
    too red and too short (i.e. much more RC-like) for its
    metallicity.}
   \label{fig:dvi}
\end{figure}

\subsection{Peculiar Globular Cluster?}

Without spectroscopic follow-up, it is not possible to judge
adequately the probability of the satellite membership for the handful
of bright blue stars in and around Crater. Leaving aside the issue of
the possible Blue Loop, how does Crater fare in comparison to the
known Galactic globulars? A different attempt to find a match for the
satellite's stellar populations is shown in
Figure~\ref{fig:cmd3}. Here, a set of Dartmouth isochrones similar to
those in Figures~\ref{fig:cmd} and ~\ref{fig:cmd2} is
over-plotted. However, this time a lower (more cluster-like) distance
modulus is assumed $(m-M)=20.8$ as well as lower $[\alpha$/Fe]=0.2
enhancement. With these adjusted parameters, the distinction between
[Fe/H]=$-$2.0 and [Fe/H]=$-$1.5 is less obvious. However, the
isochrone with [Fe/H]=$-$1.5 still lies to the red of the Crater's
RGB.

The disagreement between the model [Fe/H]=$-$1.5 population and the
Crater photometry is even more striking given how red its Horizontal
Branch is. The richness of the morphology of the Horizontal Branch in
globular clusters has yet to be explained, but it is well established
that stubby and red HBs resembling Red Clumps are a prerogative of
metal rich globulars. Note, that the Red Clump itself is rarely
invoked in the analysis of a GC as it assumes the existence of
multiple stellar populations with different ages. Metallicity is the
primary driver of the HB morphology with several other contenders for
the less important second and even third parameters. For example,
\citet{DotterHB} measure the difference in the median $V-I$ color of
the HB and the corresponding piece of the RGB for a large sample of
Galactic GCs with available HST photometry and show that age could
perhaps play the role of the second parameter, while the central
cluster density acts as a third. The study of \citet{DotterHB} shows
off beautifully the previously identified dichotomy between the inner
and outer cluster populations in the Milky Way. It appears that while
following the trend overall, at given metallicity the two sub-samples
show distinct colors and extents in their HBs: short red HBs (with low
$\Delta(V-I)$ values) typically correspond to metal-rich globulars in
the inner Galaxy, but also can be found in relatively metal-poor
systems in the outer halo.

The most conspicuous examples of this dichotomy are found in the few
most distant (i.e. beyond 80 kpc) and extended globular clusters, such
as Pal 4, 14 and AM-1. As obvious from Figure 10 of \citet{DotterHB},
the HB-RGB color difference $\Delta (V-I)$ in these reaches values as
low as 0.2. Interestingly, at the same $\Delta(V-I)$, the inner
clusters are found to be significantly more metal rich by at least 0.6
dex. Figure~\ref{fig:cmd4} compares the CMD of Crater to the stellar
populations in Pal 4 and Pal 14. Of the entire population of the
Galactic GCs, these two are perhaps the closest matches to Crater's
CMD morphology. Both are relatively metal-poor and both have short red
RHBs. However, as Figure~\ref{fig:cmd4} convincingly demonstrates
neither matches Crater's CMD perfectly. First, the Crater RGB is
clearly bluer than the RGB in either Pal 4 or Pal 14 which means that
Crater ought to be more metal-poor compared to these two. Judging by
the offset between the Crater RGB and the ridgelines of Pal 4 and 14,
our original estimate that the its metallicity is lower than -1.5, and
perhaps closer to -2.0, must not be far from the truth. Second, the
difference between the median colors of the HB and the RGB in Crater
is even smaller than in Pal 4 and 14, at
$\Delta(V-I)=0.11$. Figure~\ref{fig:dvi} is the analog of Figure 10 in
\citet{DotterHB} and aims to illustrate the unusual properties of the
HB in Crater, assuming the satellite is a globular cluster. According
to this only a handful of the most metal-rich globular clusters in the
Galaxy have a RHB situated that close to the RGB.

\begin{figure}
  \centering
  \includegraphics[width=0.99\linewidth]{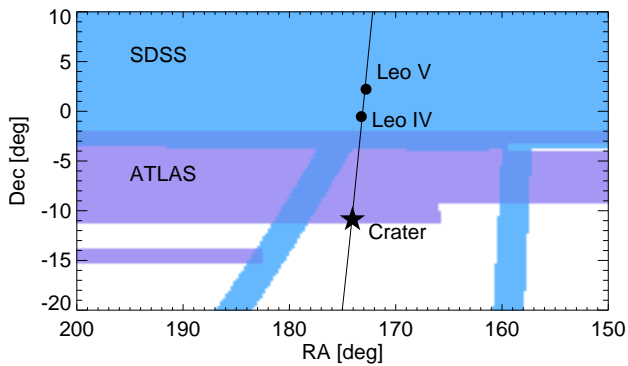}
  \caption[]{\small Large scale view of Crater's position on the
    sky. Curiously, Crater, at 170 kpc, appears to lie only $\sim
    10^{\circ}$ away from the Leo IV (155 kpc) satellite which itself is a
    neighbor of Leo V (180 kpc). The blue region gives the footprint of
    the SDSS survey in which Leo IV and V were discovered, while
    purple shows the boundaries of the current ATLAS VST footprint in 
    this region. The black line is the great circle passing very close to
    all three satellites and with a pole at
    $(\alpha,\delta)=(83.1^{\circ}, -5.3^{\circ})$.}
   \label{fig:leos}
\end{figure}

\subsection{Companions}

\begin{figure*}
  \centering
  \includegraphics[width=0.97\linewidth]{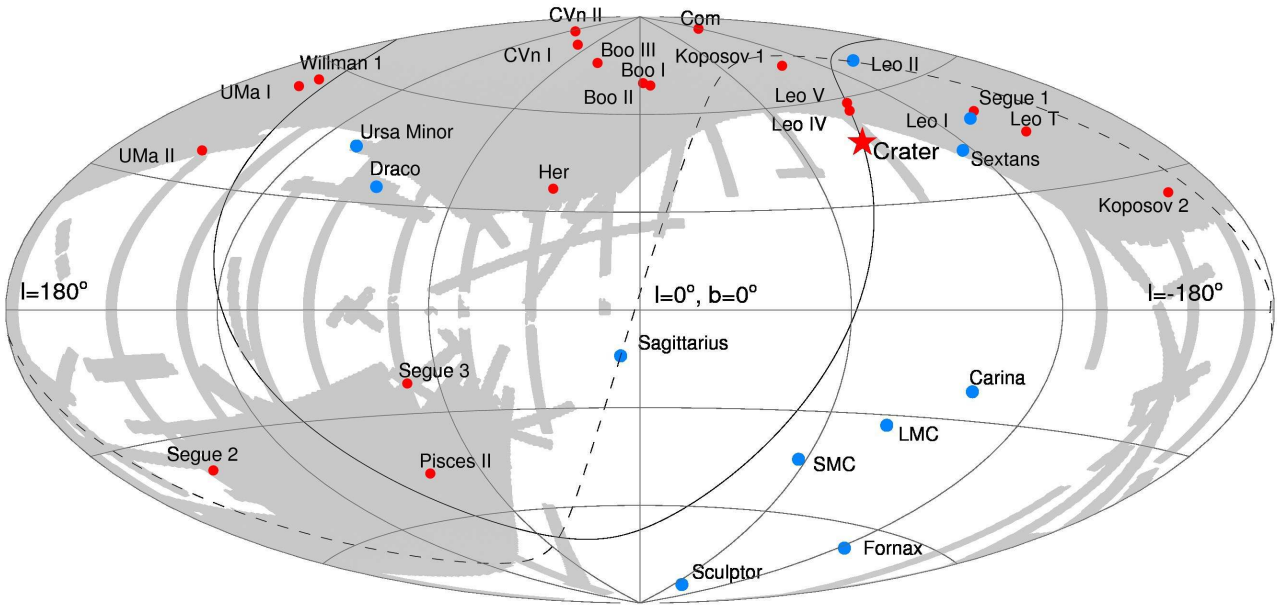}
  \caption[]{\small Positions of the classical (blue circles) and
    ultra-faint (red circles) satellites on the sky. The dashed curve
    shows the orbit of the Sgr dwarf, while the solid line represents the
    great circle with pole at $(\alpha,\delta)=(83.1^{\circ},
    -5.3^{\circ})$ passing through Crater, Leo IV and Leo V.}
   \label{fig:dwarfs}
\end{figure*}

Curiously, Crater lies only $\sim 10^{\circ}$ away from two recently
discovered ultra-faint satellite of the Galaxy: Leo IV which itself is
accompanied by Leo V some 2 degrees further to the North (see
Figure~\ref{fig:leos}). The distances to all three satellites agree
within the likely errors, therefore there seems to be an interesting
bunching of satellites over a relatively small area of sky at $160 < D
< 180$ kpc distance.  As Figure~\ref{fig:leos} illustrates, it is
possible to draw a great circle, namely with the pole at
$(\alpha,\delta)=(83.1^{\circ}, -5.3^{\circ})$, that will pass very
near Crater, Leo IV and Leo V.

De Jong et al (2010) have already found evidence for a stellar bridge
connecting Leo IV and Leo V. They show that, if the Milky Way's
potential is spherical, then Leo IV and Leo V do not share the exact
same orbit, but conclude that the galaxies are most likely an
interacting pair that fell into the Milky Way halo together. Although
the infall of groups or associations of dwarfs seems the most probable
explanation, it merits further investigation whether such an alignment
could still be purely accidental.

To complement Figure~\ref{fig:leos}, Figure~\ref{fig:dwarfs} gives the
positions of all classical and ultra-faint satellites on the sky in
Galactic coordinates. It is perhaps worth noting that the ultra-faint
satellites UMa I and Willman 1 also lie close to the great circle passing
through Crater, Leo IV and Leo V.

\section{Conclusions}
\label{sec:conclusions}

In this paper we have presented the discovery of a new Galactic
satellite found in VST ATLAS survey data. The satellite is located in
the constellation of Crater (the Cup) and has the following
properties.

\smallskip
\noindent
(1) The satellite has a half-light radius of $r_{\rm h}\sim30$ pc and
a luminosity of $M_V=-5.5$.

\smallskip
\noindent
(2) The heliocentric distance to Crater is between 145 and 170 kpc.

\smallskip
\noindent
(3) The bulk of Crater's stellar population is old and
metal-poor.

\smallskip
\noindent
(4) However, there are also several bright blue stars that appear to
be possible Crater members. We interpret these as core helium burning
Blue Loop giants and, therefore, conclude that the satellite might
have formed stars as recently as 400 Myr ago.

\smallskip
\noindent
(5) If Crater were a globular cluster than it possesses the shortest
and the reddest Horizontal Branch of all known Galactic globulars with
[Fe/H]$<-$1. Such a stubby RHB could otherwise be confused with a Red
Clump.

If our hypothesis of relatively recent star-formation is correct, then
Crater must be classified as a dwarf galaxy, as no extended
star-formation has ever been recorded in globular clusters. The
satellite then clearly stands out compared to the rest of the known
dwarfs, both classical and ultra-faint: it is much smaller and less
luminous than the classical specimens and much denser than the typical
ultra-faint ones. Given its feeble size and luminosity it is
surprising to have found (albeit tentative) evidence for a younger
population. The only other Galactic dwarf of comparable metallicity,
age and luminosity, with detected low levels of recent star-formation,
is Leo T \citep{leotdisc, RyanWeber}, which is known to have kept some
of its neutral hydrogen supply intact. This is perhaps unsurprising
given its large Galactocentric distance and very modest levels of
star-formation activity. We have attempted a search for the presence
of HI coincident with the location of Crater in the data of the GASS
radio survey \citep{GASS}, but found no convincing evidence for a
stand-alone neutral hydrogen cloud. Based on the fact that the HI gas
in Leo T is barely detected in the HIPASS data \citep{leotdisc}, and
given that Crater is significantly less luminous compared to Leo T, it
is perhaps worth pointing a radio interferometer in the direction of
this satellite.

Crater is just the most recent in a series of discoveries of ambiguous
objects that share some of the properties of clusters and dwarf
galaxies. The discoveries of Willman 1 and Segue 1 were followed by
similar controversies as to their true nature~\citep[see
  e.g.,][]{willman1disc,Ni09,Wi11,Ma11}.  Philosophers of course
recognise this as the fallacy of the excluded middle, in which a
binary choice is assumed to exhaust all the possibilities.  The
question ``Is Crater a globular cluster or a dwarf galaxy?''  might
just be as futile as the question ``Is the platypus an otter or a
duck?". The egg-laying mammal looks like both but is neither. Instead,
its mixed-up appearance is the result of it having evolved in an
unusual and isolated location.

\section*{Acknowledgments}
The authors wish to thank Emma Ryan-Weber for the expert advice
regarding the GASS data. VB has enjoyed discussing the details of this
work with Mario Mateo, Edward Olszewski, Daniel Weisz and Thomas de
Boer. The research leading to these results has received funding from
the European Research Council under the European Union's Seventh
Framework Programme (FP/2007-2013) / ERC Grant Agreement
n. 308024. Additionally, this research was made possible through the
use of the AAVSO Photometric All-Sky Survey (APASS), funded by the
Robert Martin Ayers Sciences Fund. VB acknowledges financial support
from the Royal Society. SK acknowledges financial support from the
STFC and the ERC. VB and MJI are indebted to the International Space
Science Institute (ISSI), Bern, Switzerland, for supporting and
funding the international team "First stars in dwarf galaxies".

\end{document}